\documentclass[a4paper, 12pt]{article}
\begin{document}
\title{Inferences from Bose-Einstein correlations in multiple particle production
processes
\thanks{Supported in part by the KBN grant 2P03B 093 22}
}
\author{K.Zalewski\\
M.Smoluchowski Institute of Physics,
 Jagellonian University,\\ Cracow, Reymonta 4, 30 059 Poland,\\ e-mail:
zalewski@th.if.uj.edu.pl \\ and \\ Institute of Nuclear Physics, Cracow
 }
\maketitle
\begin{abstract}
Data on Bose-Einstein correlations yield information about the interaction regions in
multiple particle production processes. The conclusions are model dependent. Several
popular models are briefly presented, compared and discussed.
\end{abstract}
Bose-Einstein correlations.\vspace{1cm}

\section{Introduction}
 An important reason for studying Bose-Einstein correlations (BEC) in multiple particle
 production processes is that they yield information about the interaction regions
 i.e. about the regions, where the hadrons are produced. There is much to learn about
 such regions. Let us consider two examples. In high energy $e^+e^-$ annihilations
 the first stages of the process can be described perturbatively. The two primary
 leptons merge into a heavy photon, or a $Z^0$, this particle decays into a
 $q\overline{q}$ pair, the quark and the antiquark radiate gluons, the gluons split
 into $q\overline{q}$ pairs or into gluons, and so on. Then, however, something much more
 complicated happens: the swarm of parton gets converted into hadrons. This is a nonperturbative
 process, known as hadronization, which is poorly understood. One would like to know
 the size and shape of the region where the hadrons are produced, the time interval
 between the first annihilation and the end of the hadronization process, as well as  the time
 interval between the beginning and the end of hadronization. As a more complicated
 example let us consider central heavy ion collisions at high energy. In this case the
 initial color fields are so strong that perturbative methods may be unreliable. It is
 usually assumed that the initial partons, whatever their production mechanism,
 rapidly thermalize. From that time on they can be considered as a fluid (a kind of
 liquid or a kind of gas), which expands according to the laws of hydrodynamics or
 according to some simplified version of the Boltzmann equation. Finally the hadrons
 are produced in a hadronization process which may be different from that in $e^+e^-$
 annihilations. Here, besides the hadronization, one is interested in the flows of the
 fluid, in its equation of state and its phase transitions, if any. The study of
 Bose-Einstein correlations among the final hadrons supplies tentative answers to all
 such questions. The problem is, however, how reliable these answers are? Till a
 few years ago the mood was optimistic (cf. e.g. \cite{WIH}). The recent results from
 RHIC, and also some results from LEP, however, have been so unexpected and so puzzling
 (references can be traced e.g. from the recent review \cite{ZAL1}) that perhaps some
 important ideas are still missing. In the present paper we review some of the most
 popular ways of getting from the experimental data to the physical conclusions. It
 will be seen that much remains to be clarified there.

 The key object in the transition from the experimental data to the physical
 conclusions is usually a single particle density matrix
 $\rho(\textbf{p}_1,\textbf{p}_2)$.
 There are approaches where this is not the case, e.g. the string model (cf. \cite{AND}
 and references quoted there), but most models can be formulated in terms of this
 matrix \cite{ZAL2}. In this section we will present its
 relation to the data. The discussion is grossly simplified (cf. e.g. \cite{WIH}), but
 it contains the main ideas. The relation between the density matrix and the features of
 the interaction region is the main subject of this paper and is discussed in the
 following sections. Let us stress that the density matrix
 $\rho(\textbf{p}_1,\textbf{p}_2)$ is an auxiliary construct, which does not have to
 coincide with the actual single particle density matrix for the mesons of a given
 type in the final state.

 The diagonal elements of the density matrix are related, as usual, to the single
 particle momentum distribution

\begin{equation}\label{}
\frac{dN}{d^3p} \sim \rho(\textbf{p},\textbf{p}).
\end{equation}
The tilde in most models means equality up to a constant factor. Only in the GGLP
model it is a more complicated operation. $d^3p$ may denote the infinitesimal volume
in momentum space, or the covariant infinitesimal volume i.e. the volume in momentum
space divided by the corresponding energy $E(\textbf{p})$. The two-particle
distribution for identical bosons is

\begin{equation}\label{}
  \frac{dN}{d^3p_1d^3p_2} \sim \rho(\textbf{p}_1,\textbf{p}_1)\rho(\textbf{p}_2,\textbf{p}_2)
  + |\rho(\textbf{p}_1,\textbf{p}_2)|^2.
\end{equation}
For uncorrelated, distinguishable particles the second term on the right-hand side
would be absent. It results from the symmetrization of the product of the single
particle density matrices of the two particles and thus reflects the BEC. Usually one
considers the correlation function obtained by dividing the two-particle distribution
by the product of the corresponding single particle distributions

\begin{equation}\label{corfun}
C(\textbf{p}_1,\textbf{p}_2) \sim 1 +
\frac{|\rho(\textbf{p}_1,\textbf{p}_2)|^2}{\rho(\textbf{p}_1,\textbf{p}_1)\rho(\textbf{p}_2,\textbf{p}_2)}.
\end{equation}
In practice the procedure is usually much more complicated than presented here, but
the strategy is as described: the data are used to get the correlation function which
is simply related to the single particle density matrix.

\section{GGLP approach}

The first model of BEC in multiple particle production was proposed by the Goldhabers
Lee and Pais \cite{GGL} (GGLP). As a realistic description of BEC in multiple particle
production processes this model is outdated, but it is still the best introduction to
the subject. GGLP introduced a density $\rho(\textbf{x})$ of particle sources. They
considered identical pions and for definiteness also we will call the identical bosons
pions, though the analysis can be applied to any identical bosons (or even to
identical fermions, when the sign between the two terms in formula (\ref{corfun}) is
changed from plus to minus). GGLP made the following two assumptions: firstly, that
all the pions are produced simultaneously and instantaneously; secondly, that the
production is completely incoherent -- the pions produced in two different space point
do not interfere. Under these assumptions the single particle density operator is

\begin{equation}\label{}
  \hat{\rho} = \int d^3x |\textbf{x}\rangle \rho(\textbf{x})\langle \textbf{x}|.
\end{equation}
The corresponding density matrix reads

\begin{equation}\label{gglrho}
  \rho(\textbf{p}_1,\textbf{p}_2) =
  \langle \textbf{p}_1|\hat{\rho}|\textbf{p}_2 \rangle \sim
  \int d^3x e^{-i\textbf{qx}}\rho(\textbf{x}),
\end{equation}
where $\textbf{q} = \textbf{p}_1 - \textbf{p}_2$. If this were the density matrix
which can be obtained from formula (\ref{corfun}), it would be a very nice result. The
density of sources $\rho(\textbf{x})$ could then be obtained unambiguously just by
inverting the Fourier transform. E.g. for

\begin{equation}\label{}
  \rho(p_1,p_2) \sim e^{-\frac{1}{2}R^2 \textbf{q}^2},
\end{equation}
where $R^2 >0$ is a constant, one would obtain

\begin{equation}\label{}
  \rho(\textbf{x}) \sim e^{-\frac{\textbf{x}^2}{2R^2}}.
\end{equation}
Unfortunately, the density matrix (\ref{gglrho}) is untenable as a density matrix
proportional to the true single particle density matrix, because it gives a single
particle momentum distribution constant in all momentum space. GGLP introduced,
therefore, a projection on the states allowed by energy-momentum conservation, i.e.
they used the density matrix as calculated here to find the momentum distribution for
all the particles present in the final state, multiplied it by the delta function of
energy-momentum conservation and integrated over the (covariant) momentum space of all
the particles except two identical pions for the two particle distribution and over
all the momenta except one for the single particle distribution. They got results in
qualitative agreement with experiment. The method, however, was cumbersome -- no one
did calculations for more than six particles in the final state -- and was in violent
disagreement with experiment at the quantitative level \cite{CZS}.

\section{Kopylov and Podgoretskii model}

Very significant progress was obtained by Kopylov and Podgoretskii (cf. \cite{KOP} and
references contained there). In their model the density operator in the Schr\"odinger
picture is

\begin{equation}\label{}
  \hat{\rho} = \int d^4x_s e^{iH_0(t_s - t)}|\psi_s\rangle\rho(x_s)\langle \psi_s|
  e^{-H_0(t_s - t)},
\end{equation}
where

\begin{equation}\label{}
  \langle \textbf{x}|\psi_s\rangle = \psi(\textbf{x} - x_s);
\end{equation}
or equivalently

\begin{equation}\label{}
  \langle \textbf{p}|\psi_s\rangle \sim \int d^3x e^{-i\textbf{px}}\psi(x - x_s)
  = e^{-i\textbf{px}_s}A(\textbf{p}),
\end{equation}
where

\begin{equation}\label{}
  A(\textbf{p}) = \int d^3x e^{-i\textbf{px}}\psi(\textbf{x}).
\end{equation}
In these formulae $x_s$ is a point in space-time, as well as a label which defines
unambiguously a source. It should be related to the position of the corresponding
source in space-time, but the actual relation is a matter of choice. E.g. it could be
the point where the source got created, or its mean position in space-time. Each
source is labelled by a space-time point, but the particle produced by a source is not
localized in a point. Its wave function in space is $\psi(\textbf{x}-\textbf{x}_s)$
with an additional time dependent phase which is zero at $t = t_s$. All these
functions are related by shifts in space-time. The wave function in momentum space is
proportional to $A(\textbf{p})$, thus all the sources yields particles with the same
momentum distribution. $H_0$ is the free particle Hamiltonian and we are interested in
times $t$ larger than the latest time $t_s$, thus the particle evolves freely.

The corresponding density matrix,

\begin{equation}\label{denskp}
  \rho(\textbf{p}_1,\textbf{p}_2) \sim A(\textbf{p}_1)A^*(\textbf{p}_2)
  \int d^4x_x e^{iqs}\rho(x_s),
\end{equation}
yields the single particle distribution,

\begin{equation}\label{}
  \frac{dN}{d^3p} \sim |A(\textbf{p})|^2,
\end{equation}
which can be made to agree with any experimental distribution by a suitable choice of
$A(\textbf{p})$. Thus, there is no obvious need to introduce the energy-momentum
conservation constraint. Kopylov and Podgoretskii assumed that for final states with
not too few particles in the interesting momentum region the energy-momentum
conservation does not affect significantly the one- and two-body distributions, and
that consequently their matrix (\ref{denskp}) can be assumed equal, up to a
normalizing factor, to the actual single particle density matrix. This assumption
makes it easy to get results for high multiplicity exclusive channels and for high
energy inclusive processes. Both were beyond the reach of the GGLP method. Besides the
satisfactory formula for the single particle distribution given above, the model gives
for the correlation function

\begin{equation}\label{}
  C(\textbf{p}_1,\textbf{p}_2) = 1 + \int d^4x_s e^{iqx_s} \rho(x_s).
\end{equation}
Here the factors $A(p)$ cancel and the result is as for point sources. Note, however,
that because of the four-fold integration this relation cannot be inverted to give
$\rho(x_s)$, when the correlation function is known. Moreover, a priori the
correlation function could depend on the vector $\textbf{q}$ and on the vector
$\textbf{K} = (\textbf{p}_1 + \textbf{p}_2)/2$, while the right hand side depends only
on the four-vector $q$. The reason is that the factor $A(\textbf{p})$ is the same for
every source and cancels. Physically this means that there is no correlation between
the momentum of the pion and the position of the source. Since such correlation follow
from almost every model and since experimentally it is not true, that the correlation
function depends only on $q$ (cf. e.g. \cite{WIH} and references quoted there) this is
a serious weakness of the model.

\section{Yano and Koonin method}

Another attempt to go beyond the GGLP approximations is due to Yano and Koonin
\cite{YAK}. In our notation their key formula is

\begin{equation}\label{}
  |\rho(\textbf{p}_1,\textbf{p}_2)|^2 = Re \int d^4x_1 D(x_1,\textbf{p}_1)
  e^{iqx_1}\int d^4x_2 D(x_2,\textbf{p}_2)e^{-iqx_2}.
\end{equation}
Putting

\begin{equation}\label{}
  D(x,\textbf{p}) = \delta(t)\rho(\textbf{x})
\end{equation}
one recovers the GGLP model. Putting

\begin{equation}\label{jankoo}
  D(x,\textbf{p}) = |A(\textbf{p})|^2\rho(x),
\end{equation}
which corresponds to the assumptions made in \cite{YAK}, one reproduces the results of
Kopylov and Podgoretskii. It is not clear, whether this approach can be given sense in
the framework of quantum mechanics, when there are position-momentum correlations,
i.e. when the $D$ function does not factorize into a momentum dependent and a space
dependent factor.

\section{Covariant current formalism}

The covariant current formalism (cf. \cite{PGG} and references given there) can be
derived from the model of Kopylov and Podgoretskii introducing the following two
generalizations. The label characterizing the source is changed from $x_s$ to
$x_s,p_s$. Thus, the source is characterized by its position in space-time and by its
four momentum. Correspondingly, the density of sources $\rho(x_s)$ gets replaced by
$\rho(x_s,p_s)$. Such labels are not subject to the Heisenberg uncertainty principle.
E.g. a one-dimensional harmonic oscillator may have $\langle  x \rangle = 0$ and
$\langle p \rangle
 = 0$. The universal, momentum dependent function $A(\textbf{p})$ gets replaced by the
source dependent $j(\frac{p_sp}{m_s})$, where $m_s$ is the mass and $p_s$ the momentum
of the source. In the Kopylov Podgoretskii model all the sources had the same momentum
distribution in some overall reference frame, e.g. in the center of mass frame of the
collision. In the covariant current formalism each source has the same momentum
distribution, when considered \textit{in its rest frame}. Another way of formulating
this model is to replace the incoherent sources by incoherent wave packets. Then $p_s$
and $x_s$ characterize the wave packet, and since the waves in the packet describe
pions, it is natural to put $m_s = m_\pi$. In this model, assuming a fixed mass for
all the sources,

\begin{equation}\label{}
  \rho(p_1,p_2) = \int d^4x_s \int d^3p_s \rho(x_s,p_s)e^{iqs_s}
  j(\frac{p_sp_1}{m_s})j^*(\frac{p_sp_2}{m_s}).
\end{equation}
From the point of view of model builders, a nice feature of this approach is that one
can assume a classical motion of the source, e.g. $\textbf{x}_s = \textbf{x}_s(t_s)$
and $\textbf{p}_s = \textbf{p}_s(t_s)$, and still have a formula which is consistent
with quantum mechanics. In the covariant current formalism position-momentum
correlations are naturally included.

\section{Emission function method}

The method of emission functions is very popular nowadays (cf. e.g. \cite{WIH}). The
emission function is built by analogy with the Wigner function \cite{PRA1},
\cite{PRA2}. The Wigner function $W$ is related to the single particle density matrix
in the momentum representation by the formula

\begin{equation}\label{}
  \rho(\textbf{p}_1,\textbf{p}_2) = \int d^3X e^{-i\textbf{qX}} W(\textbf{X},\textbf{K}),
\end{equation}
Where $\textbf{X} = (\textbf{x}_1 + \textbf{x}_2)/2$, and as usual $\textbf{q} =
\textbf{p}_1 - \textbf{p}_2$ and $\textbf{K} = (\textbf{p}_1 + \textbf{p}_2)/2$. The
Wigner function is well-defined as a Fourier transform of the density matrix in
momentum representation. On the other hand, as a function of $\textbf{X}$ and
$\textbf{K}$ it may play the role of a phase space distribution. When studying the
interactions regions, it is important to combine information about the space and the
momentum distributions. Kopylov, Podgoretskii, and followers proposed to use the
position of the source and the momentum of the particle. These two vectors can be
measured simultaneously and their joint distribution gives an idea about the phase
space distribution of particles. The Wigner function gives another ersatz phase space
distribution. The trouble is, however, that the Wigner function refers to a given
moment of time, while the hadrons are produced during a time interval. The emission
function $S$ is supposed to improve on that and is related to density matrix by the
formula

\begin{equation}\label{}
  \rho(\textbf{p}_1,\textbf{p}_2) = \int d^4X e^{iqX}S(X,K).
\end{equation}
The usual strategy is to find from some classical or quasi-classical argument a phase
space distribution as function of time, and then to interpret it as the emission
function. Once this is done, one can, by a four-fold integration, obtain the density
matrix and compare it with experiment. This comparison may be used to fix the free
parameters of the model.

Let us note some difficulties of this approach.  Given a density matrix there is an
infinity of very different emission functions, which can reproduce it. For instance
perfect agreement is obtained for

\begin{equation}\label{}
  S(X,K) = \delta(X_0)W(\textbf{X},\textbf{K}).
\end{equation}
This, however, corresponds to simultaneous and instant production of all the hadrons,
which is not a very plausible scenario. When particle production is an incoherent sum
of production amplitudes at various moments of time, the emission function
$S(\textbf{X},t,K)$ can be related to the Wigner function of the particles produced at
time $t$. In the general case, however, the relation of the emission function to a
Wigner function is hardly visible \cite{ZAL3}. Thus, the formula can be used to
eliminate wrong models rather than to prove that a model is implied by the data. This
is not necessarily bad. Eliminating the unacceptable values of the parameters of a
model, one learns which are the good ones. Nevertheless, one must always keep in mind
that a completely different model can give equally good, or better results. One should
also keep in mind that $\textbf{X}$ and $\textbf{K}$ are not really position and
momentum, but only half sums of the corresponding arguments of the density matrix.
When the contributions of the sources are strongly smeared in coordinate space or in
momentum space, the difference may be very significant.


\begin{thebibliography} {99}
\bibitem{WIH}U.A. Wiedemann, U. Heinz, Phys. Rep. 319,145(1999).
\bibitem{ZAL1}K. Zalewski, Surprises in Bose-Einstein correlation, report at QCD'03
Montpellier July 2003 and hep-ph.
\bibitem{AND}B. Andersson, Acta Phys. Pol. B29,1885(1998).
\bibitem{ZAL2}K. Zalewski, Acta Phys. Pol. B33,2643(2003).
\bibitem{GGL}G. Goldhaber, S. Goldhaber, W. Lee, A. Pais, Phys. Rev. 120,300(1960).
\bibitem{CZS}O. Czy\.{z}ewski, M. Szeptycka, Phys. Lett. 25B,482(1967).
\bibitem{KOP}G.I. Kopylov, M.I. Podgoretskii, Yad. Phys. 19,434(1974).
\bibitem{YAK}F.B. Yano, S.E. Koonin, Phys. Lett. B78,556(1978).
\bibitem{PGG}S.S. Padula, M. Gyulassy, S. Gavin, Nucl. Phys. B329,357(1990).
\bibitem{PRA1}S. Pratt, Phys. Rev. Lett. 53,1219(1984).
\bibitem{PRA2}S. Pratt, Phys. Rev. D33,72(1986).
\bibitem{ZAL3}K. Zalewski, Acta Phys. Pol. B34,3379(2003).
\end{thebibliography}
\end{document}